\documentclass[aps,amssymb,floatfix,prd,amsmath,preprintnumbers]{revtex4}
\usepackage{graphicx}  
\usepackage{float}
\usepackage{dcolumn}   
\usepackage{bm}
\begin{document}

\title{\bf  Accelerating models with a hybrid scale factor in extended gravity}

\author{B. Mishra \footnote{Department of Mathematics, Birla Institute of Technology and Science-Pilani, Hyderabad Campus, Hyderabad-500078, India, E-mail:bivudutta@yahoo.com },  S. K. Tripathy\footnote{Department of Physics, Indira Gandhi Institute of Technology, Sarang, Dhenkanal, Odisha-759146, India, E-mail:tripathy\_ sunil@rediffmail.com}, Sankarsan Tarai \footnote{Department of Mathematics, Birla Institute of Technology and Science-Pilani, Hyderabad Campus, Hyderabad-500078, India, E-mail:tsankarsan87@gmail.com}
}\affiliation{ }

\begin{abstract}
Dynamical aspects of cosmological model in an extended gravity theory have been investigated in the present work. We have adopted a simplified approach to obtain cosmic features, which in fact requires more involved calculations. A cosmological model is constructed using a hybrid scale factor that simulates a cosmic transit behaviour.  The deceleration parameter and energy conditions have been obtained for the constructed model. Scalar fields have been reconstructed from the present model in the extended gravity. Different diagnostic methods have been applied to analyse the viability of the constructed model. The present model almost looks like a cosmological constant for a substantial cosmic time zone and does not show any slowing down feature in near future. 
\end{abstract}
\maketitle
\section{Introduction} 
The astronomical observations revealed two important phenomena on the universe: the dark energy and the dark matter. Dark energy yields a late time acceleration of the cosmological background whereas dark matter behaves as an invisible dust matter favouring the process of gravitational clustering. The concept of the intriguing matter such as dark energy has emerged as an alternative to the General Relativity (GR) based description of gravity \cite{Tsujikawa10, Capozziello11, Nojiri11, Clifton12, Berti15}. The search for alternative theories to GR has become inevitable after its failure to explain the cosmic phenomena occurring at late times. GR can be modified in many distinct directions. Consequently there are a good number of modified gravity theories available in  literature \cite{Capozziello03, Nojiri03, Carrol04, Sotiriou10,  Nojiri11}. Different exotic matter fields simulating a positive energy density and negative  pressure  are commonly used in literature as additional terms to explain the cosmic speed up phenomenon. However, it is possible to settle the issue of late time cosmic acceleration through the suitable geometrical modification in GR without adding any exotic source of matter. The most simple geometrical modification to GR has been proposed is the $f(R)$ theory, where the geometrical action contains an arbitrary function $f(R)$ of the Ricci scalar $R$ in place of the simple $R$ in GR action. Some generalizations of $f(R)$ gravity have been proposed recently. One such generalization is the $f(R,T)$ gravity proposed by Harko et al. \cite{Harko11}. In $f(R,T)$ gravity, a weak coupling between the geometry and matter is assumed and accordingly the geometrical part of the gravitational field action is modified by considering an arbitrary function $f(R,T)$ of the Ricci scalar $R$ and the trace of the energy momentum tensor $T$.   It is worthy to mention here that employing the trace of the energy momentum tensor in the new theory may be associated with the existence of exotic imperfect fluids or quantum effects such as particle production.

Many researchers have studied different aspects of $f(R,T)$ gravity in various physical background. Sharif and Zubair \cite{Sharif14} have studied the cosmological reconstruction of $f(R,T)$ gravity in FRW universe. Shabani and Farhoudi \cite{Shabani14} obtained the cosmological parameters in terms of some defined dimensionless parameters that are used in constructing the dynamical equations of motion. Dynamics of an anisotropic universe is studied by Mishra et al. \cite{Mishra16} in $f(R,T)$ gravity using a rescaled functional whereas Yousaf et al. \cite{Yousaf16} have shown the causes of irregular energy density in $f(R,T)$ gravity. Velten and Carames \cite{Velten17} have challenged the viability of $f(R,T)$ as an alternative modification of gravity. Abbas and Ahmed \cite{Abbas17} have formulated the exact solutions of the non-static anisotropic gravitating source in $f(R,T)$ gravity which may lead to expansion and collapse. Baffoul et al. \cite{Baffoul17} have investigated the late-time cosmic acceleration in mimetic $f(R,T)$ gravity with the Lagrange multiplier. Carvalho et al. \cite{Carvalho17} have shown the equilibrium configurations of white dwarfs in a modified gravity theory. Mishra et al. \cite{Mishra18a} developed a general formalism to investigate Bianchi type $VI_h$ universes in an extended theory of gravity whereas the the dynamical features of the models have been studied in \cite{Mishra18b, Mishra18c}. The investigation of different aspects of the cosmic phenomena concerning the late time dynamics including the isotropic and anisotropic nature in the modified gravity theory requires an involved calculations of the dynamical parameters. Also, the results of the calculations should be in conformity with the lot of information gathered over the years from observations. In view of this, in the present work, we have developed a formalism to investigate the dynamical cosmic features in the framework of $f(R,T)$ gravity. We have considered a Bianchi type anisotropic and homogeneous universe for this purpose.

The paper is organised as follow: in Section II, the basic formalism of $f(R,T)$ gravity and the field equations for Bianchi type $VI_h$ space time have been derived. In Section III, anisotropic nature of the cosmological model has been presented. The dynamical parameters of the model have been calculated and analysed in Section IV. A scalar field reconstruction technique has been employed in Section V to obtain the behaviour of the reconstructed scalar fields. Keeping in view the dynamically varying nature of the deceleration parameter in passing from a decelerated phase of expansion to an accelerated one, we have employed a hybrid scale factor (HSF) in the proposed formalism to investigate the late time cosmic dynamics in Section VI. The viability of the constructed models have tested through different diagnostic mechanisms in Section VII  and finally the concluding remarks are given in section VIII.

\section{Basic Formalism of $f(R,T)$ theory and the field equations for a $BVI_h$ space-time}
Within the scope of an extended gravity theory as proposed by Harko et al.\cite{Harko11}, the Einstein-Hilbert action is given by
\begin{equation} \label{eq:1}
S=\int d^4x\sqrt{-g}\left[\frac{1}{16\pi} f(R,T)+ \mathcal{L}_m \right].
\end{equation}
In the above, we have used the natural unit system  so that $G=c=1$, where $G$ and $c$ are respectively the Newtonian gravitational constant and speed of light in vacuum. $\mathcal{L}_m$ is the matter Lagrangian. The above action is different from that of GR where an arbitrary function of $R$ and $T$ ($f(R,T)$) replaces the Ricci scalar $R$.  This interesting coupling of matter and curvature is motivated from quantum effects and leads to a non vanishing divergence of the energy-momentum tensor $T_{ij}$. Such a feature of the non-minimal matter-geometry coupling provides a strong ground for cosmic acceleration. Because of this coupling, an additional force comes into play that dispels massive particles away from geodesic trajectories. Since the functional form governing the matter-curvature coupling is arbitrary, different choices of the functional $f(R,T)$ generate different models. In general, there can be three possible functional ways of coupling namely: (i) $f(R,T)=R+2f(T)$, (ii) $f(R,T)=f_1(R)+f_2(T)$ and (iii) $f(R,T)=f_1(R)+f_2(R)f_3(T)$, where $f(T), f_1(R), f_2(R), f_2(T)$ and $f_3(T)$ are some arbitrary functions of their respective arguments. In fact, there can be infinite number of ways to chose these functions. However, the viability of the constructed models depends on the suitable choice of these functions and their ability to pass certain geometrical and observational tests. Out of these three types, the first choice is more like GR and can be reduced to GR under certain condition. The second term of the first choice in that case can be considered as a small deviation from GR. In the present work, we consider the functional as $f(R,T)=R+2\Lambda_0+2\beta T$ where, $\Lambda_0$ is the usual time independent cosmological constant and $\beta$ is a coupling constant. This functional form resembles the first type of coupling as mentioned with $R$ being replaced by $R+2\Lambda_0$. The GR features with a cosmological constant can be recovered from the model for a vanishing $\beta$. It is needless to mention here that the choice of the functional of the type $f(R,T)=R+2f(T)$ has been widely used in literature to address different cosmological and astrophysical issues. 

The field equations for the present model can be obtained by varying the action. One can refer to the procedure followed to obtain the field equation in Refs \cite{Harko11, Mishra16}. For an arbitrary choice  $f(R,T)=f(R)+f(T)$ and $\mathcal{L}_m=-p$, the field equations are obtained as

\begin{equation} \label{eq:2}
R_{\mu\nu}-\frac{1}{2}\frac{f(R)}{f_R (R)}g_{\mu\nu}=\frac{1}{f_R (R)} \left[\left(\nabla_{\mu} \nabla_{\nu}-g_{\mu\nu}\Box\right)f_R(R)+\left[8\pi +f_T(T)\right]T_{\mu\nu}+\left[f_T(T)p+\frac{1}{2}f(T)\right]g_{\mu\nu}\right].
\end{equation}
In the above, $p$ is the pressure of cosmic fluid. The partial differentiations for the model become $f_R=\frac{\partial f(R)}{\partial R} =1$ and $f_T=\frac{\partial f(T)}{\partial T}=2\beta$, so that we can express the field equations as
\begin{equation} \label{eq:3}
G_{\mu\nu}= \kappa T_{\mu\nu}^{eff},
\end{equation}
with
\begin{eqnarray} \label{eq:3a}
G_{\mu\nu}&=& R_{\mu\nu}-\frac{1}{2}Rg_{\mu\nu},\\
T_{\mu\nu}^{eff} &=& T_{\mu\nu} + \frac{\Lambda_{eff}(T)}{8\pi+2\beta}~ g_{\mu\nu}.
\end{eqnarray}
Here, we have a redefined Einstein matter-geometry coupling constant $\kappa=8\pi+2\beta$ and $\Lambda_{eff}(T)=\left(2p+T\right)\beta+\Lambda_0$. For $\beta=0$, $\Lambda_{eff}(T)$ becomes the usual cosmological constant $\Lambda_0$ and the above field equation reduces to the Einstein field equations with a cosmological constant. For non vanishing value of $\beta$,  $\Lambda_{eff}(T)$ becomes a time dependent quantity. 

We consider the universe to be filled with a cloud of one dimensional cosmic strings with string tension density $\xi$ aligned along the $x$-axis. The energy-momentum tensor for such a fluid is given by
\begin{equation}\label{eq:5}
T_{\mu\nu}=(p+\rho)u_{\mu}u_{\nu} - pg_{\mu\nu}-\xi x_{\mu}x_{\nu},
\end{equation}
with
\begin{equation}\label{eq:6}
u^{\mu}u_{\mu}=-x^{\mu}x_{\mu}=1
\end{equation}
and 
\begin{equation}\label{eq:7}
u^{\mu}x_{\mu}=0.
\end{equation}
 
$\rho$ represents the energy density and is composed of the particle energy density $\rho_p$ and the string tension density $\xi$, $\rho=\rho_p+\xi$.

 We wish to investigate dynamical aspects of the universe in $f(R,T)$ theory as described above for an anisotropic Bianchi type $VI_h$ ($BVI_h$) space-time given by 

\begin{equation}\label{eq:4}
ds^2 = dt^2 - A^2dx^2- B^2e^{2x}dy^2 - C^2e^{2hx}dz^2,
\end{equation}
where the metric potentials are considered only to depend on cosmic time. As has already been discussed in some works of Tripathy et al.\cite{skt15} and Mishra et al.\cite{skt16}, the exponent $h$ in the metric can be useful if it assumes the value $h=-1$. This conclusion has been derived from the null total energy concept of the whole universe known as the Tryon's conjecture \cite{skt15, skt16}. Going along the same line of thought, in the present study, we consider the same value of $h$. Now, the field equations for $BVI_h$ space-time with $h=-1$ in the extended theory of gravity can be written as
\begin{equation} \label{eq:8}
\frac{\ddot{B}}{B}+\frac{\ddot{C}}{C}+\frac{\dot{B}\dot{C}}{BC}+ \frac{1}{A^2}= -\alpha(p-\xi) +\rho \beta+\Lambda_0,   
\end{equation}
\begin{equation} \label{eq:9}
\frac{\ddot{A}}{A}+\frac{\ddot{C}}{C}+\frac{\dot{A}\dot{C}}{AC}- \frac{1}{A^2}=-\alpha p +(\rho+\xi)\beta+\Lambda_0,   
\end{equation}
\begin{equation} \label{eq:10}
\frac{\ddot{A}}{A}+\frac{\ddot{B}}{B}+\frac{\dot{A}\dot{B}}{AB}- \frac{1}{A^2}=-\alpha p +(\rho+\xi)\beta+\Lambda_0,  
\end{equation}
\begin{equation} \label{eq:11}
\frac{\dot{A}\dot{B}}{AB}+\frac{\dot{B}\dot{C}}{BC}+\frac{\dot{C}\dot{A}}{CA}-\frac{1}{A^2}=
\alpha \rho -\left(p-\xi\right)\beta +\Lambda_0,   
\end{equation}
\begin{equation} \label{eq:12}
\frac{\dot{B}}{B}=\frac{\dot{C}}{C}.
\end{equation} 
 
Here $\alpha=8\pi+3\beta$ and we denote the ordinary time derivatives as overhead dots. 

Some relevant quantities in the context of discussion of geometrical aspect of the model include
\begin{eqnarray}
\text{Hubble rate:}~~~ H &=& \frac{\dot{\mathcal{R}}}{\mathcal{R}}=\frac{1}{3}\left(\frac{\dot{A}}{A}+2\frac{\dot{B}}{B}\right),\label{eq:16}\\
\text{Expansion scalar:}~~~ \theta &=& u_{;l}^l=\left(\frac{\dot{A}}{A}+2\frac{\dot{B}}{B}\right),\label{eq:17}\\
\text{Deceleration parameter:}~~~ q &=&  -1+\frac{d}{dt}\left(\frac{1}{H}\right).\label{eq:20}
\end{eqnarray}

\section{Anisotropic nature of the model}
In the present work, we have considered a spatially homogeneous and anisotropic $BVI_h$ universe with different expansion rates along different spatial directions. The quantities that measure the departure from spatial isotropy are the Shear scalar $\sigma^2$ and the average anisotropy parameter $\mathcal{A}$ defined respectively as 
\begin{eqnarray}
\sigma^2 &=& \frac{1}{2}\left(\sum H_i^2-\frac{1}{3}\theta^2\right)= \frac{1}{3}\left(\frac{\dot{A}}{A}-\frac{\dot{B}}{B}\right)^2,\label{eq:18}\\
\mathcal{A} &=& \frac{1}{3} \sum_{i=1}^3\left(\frac{\triangle H_i}{H}\right)^2.\label{eq:19}
\end{eqnarray}
$\triangle H_i=H_i-H$, where $H_1=\frac{\dot{A}}{A}$, $H_2=\frac{\dot{B}}{B}$ and $H_3=\frac{\dot{C}}{C}$ are the directional Hubble rates along x, y and z-axes respectively. In view of eq.\eqref{eq:12}, we have $H_2=H_3$. For isotropic models these quantities $\sigma^2$ and $\mathcal{A}$ identically vanish. From observational perspectives, the anisotropic nature of a model is usually quantified through the estimation of the amplitude of shear $\frac{\sigma}{H}$ at the present epoch. Using the data from differential microwave radiometers aboard the Cosmic Background Explorer (COBE), Bunn et al. have placed an upper limit to this quantity as $\left(\frac{\sigma}{H}\right)_0 < 3 \times 10^{-9}$ \citep{Bunn96}. For the best case with $\Omega_0=1$, they have obtained $\left(\frac{\sigma}{H}\right)_{pl} \simeq 10^{-3}- 10^{-4}$ \citep{Bunn96}. In that work Bunn et al. have concluded that primordial anisotropy should have been fine tuned to be less than $10^{-3}$ of its natural value in the Planck era. Saadeh et al. used cosmic microwave background temperature and polarisation data from Planck and  obtained a tighter limit to the anisotropic expansion as  $\left(\frac{\sigma}{H}\right)_0 < 4.7 \times 10^{-11}$ \citep{Saadeh16}. In view of these recent observational limits on cosmic shear and anisotropic expansion rates, we have adopted a simple approach in the present work and have assumed a proportional relationship between the amplitude of shear $\sigma$ and Hubble rate. This assumption leads to an anisotropic relation among the directional Hubble rates $H_1=kH_2$. The parameter $k$ takes care of the anisotropic feature of the model. Obviously, $k \neq 1$ provides an anisotropic model.

For the $BVI_h$ metric, we obtain the amplitude of shear expansion in the present epoch as
\begin{eqnarray}
\left(\frac{\sigma}{H}\right)_0=\sqrt{3}\left(\frac{k-1}{k+2}\right).
\end{eqnarray}
Even though tighter constraints on cosmic anisotropy are available in literature and evidences against the departure from global isotropy are being gathered \citep{Deng18}, these observational analysis need to be fine tuned as the analysis are prior dependent \citep{Saadeh16a}. It is worth to mention here that, the cosmological principles assuming a homogeneous and isotropic universe may be a good approximation to the present universe. However, it has not yet been well proven in the scales $\geq 1 Gpc$ \cite{Caldwell2008}. In view of this, in the present work, we wish to construct some accelerating anisotropic models keeping enough room for any amount of cosmic anisotropy.  However, we can set some constraints on the parameter $k$ basing upon the observationally found upper bound on $\left(\frac{\sigma}{H}\right)_0$. While the bounds of Bunn et al. constrain $k$ as $k=1.000000008$, that of Saadeh et al. disfavours any classical finite departure from $k=1$. However, in the present work, we consider $k=1.0000814$ that provides $\left(\frac{\sigma}{H}\right)_0 = 4.7 \times 10^{-5}$.

Within the formalism discussed here it is easy to show that $\mathcal{A}= \frac{2}{9}\left(\frac{\sigma}{H}\right)^2$. Consequently, the average anisotropic parameter in the present epoch can be calculated as $\mathcal{A}_0= 4.91\times 10^{-10}$. Since the large scale structure of the universe may show a departure from isotropy, the cosmic anisotropy can be estimated from Hemispherical asymmetries in the Hubble expansion. In a recent work, Kalus et al. estimated the Hubble anisotropy of supernova type Ia Hubble diagrams at low redshifts ($z < 0.2$) as $\frac{\triangle H}{H} < 0.038$ \citep{Kalus11}. Using the value of the anisotropy parameter $k$ at the present epoch, we obtain the expansion asymmetry as $\frac{\triangle H}{H} = 0.814\times 10^{-4}$. One can note that, the predicted anisotropy from our model is well within the observationally set up bounds \citep{Campa11}.

\section{Equation of State parameter and Energy Conditions}
The presumed anisotropic relation among the directional Hubble rates has a simplified structure and within this formalism it can provide us a simple approach to study the cosmic dynamics. For a given anisotropic parameter $k$, the directional Hubble rates become $H_1=\left(\frac{3k}{k+2}\right)H$ and $H_2=H_3=\left(\frac{3}{k+2}\right)H$. Obviously for $k=1$, the directional Hubble rates become equal to the Hubble parameter $H$. In our formalism, the presumption of the proportionality relation between the shear scalar and scalar expansion leads to the calculation of the equation of state (EoS) parameter in terms of the Hubble rate. Also, the energy conditions for the present model will depend on the Hubble rate. 

\subsection{EoS parameter}
The physical properties of the model such as pressure, energy density and string tension density are obtained from the field equations \eqref{eq:8}-\eqref{eq:12} as
\begin{eqnarray}
p &=& \left(\frac{1}{\alpha^2-\beta^2}\right)\left[\left(S_1(H)-S_2(H)+S_3(H)\right)\beta-S_2(H)\alpha+\left(\alpha-\beta\right)\Lambda_0\right],\label{eq:19}\\
\rho &=& \left(\frac{1}{\alpha^2-\beta^2}\right)\left[S_3(H)\alpha-S_1(H)\beta-\left(\alpha-\beta\right)\Lambda_0\right], \label{eq:20}\\
\xi &=& \frac{S_1(H)-S_2(H)}{\alpha-\beta}, \label{eq:21}
\end{eqnarray}

where 
\begin{eqnarray}
S_1(H)&=& \frac{1}{(k+2)^2}\left[6(k+2)\dot{H}+27H^2+(k+2)^2 ~ \mathcal{R}^{-\left(\frac{6k}{k+2}\right)}\right],\\
S_2(H)&=& \frac{1}{(k+2)^2}\left[3(k^2+3k+2)\dot{H}+9(k^2+k+1)H^2-(k+2)^2~ \mathcal{R}^{-\left(\frac{6k}{k+2}\right)}\right],\\
S_3(H)&=& \frac{1}{(k+2)^2}\left[9(2k+1)H^2-(k+2)^2~ \mathcal{R}^{-\left(\frac{6k}{k+2}\right)}\right].
\end{eqnarray}

Algebraic simplification of the above expressions yield 
\begin{eqnarray} 
p &=& -\left(\frac{1}{\alpha^2-\beta^2}\right)\left[\phi_1(k,\beta)\dot{H}+\phi_2(k,\beta)H^2-(\alpha+\beta)~ \mathcal{R}^{-\left(\frac{6k}{k+2}\right)}-(\alpha-\beta)\Lambda_0\right],\\ \label{eq:24}
\rho &=& \left(\frac{1}{\alpha^2-\beta^2}\right)\left[\phi_3(k,\beta)\dot{H}+\phi_4(k,\beta)H^2-(\alpha+\beta)~ \mathcal{R}^{-\left(\frac{6k}{k+2}\right)}-(\alpha-\beta)\Lambda_0\right],\\\label{eq:25}
\xi &=& \left(\frac{1}{\alpha-\beta}\right) \left[\phi_5(k)\left(\dot{H}+3H^2\right)+2~\mathcal{R}^{-\left(\frac{6k}{k+2}\right)}\right],
\end{eqnarray}
where 
\begin{eqnarray}
\phi_1(k,\beta) &=& \frac{3}{k+2}[(k+1)\alpha+(k-1)\beta],\\
\phi_2(k,\beta) &=& \left(\frac{3}{k+2}\right)^2[(k^2+k+1)\alpha+(k^2-k-3)\beta],\\
\phi_3(k,\beta) &=& -\frac{6\beta}{k+2},\\
\phi_4(k,\beta) &=& \left(\frac{3}{k+2}\right)^2[(2k+1)\alpha-3\beta],\\
\phi_5(k) &=& \frac{3(1-k)}{k+2}.
\end{eqnarray}

It is interesting to note that for $\alpha+\beta=0$ i.e. for $\beta=-2\pi$ we have 
\begin{equation}
\phi_1(k,\beta)=\phi_3(k,\beta)~~~~ \text{and}~~~~~ \phi_2(k,\beta)=\phi_4(k,\beta)
\end{equation}
and consequently in the limit $\beta\rightarrow -2\pi$,
\begin{equation}
p=-\rho.
\end{equation}
In other words, within the scope of the present formalism, $\Lambda$CDM model with $p=-\rho$ can be recovered from the model for $\alpha+\beta=0$. Of course, overlapping of the present model with that of $\Lambda$CDM requires a negative coupling constant.

The equation of state parameter (EoS) $\omega$ is defined as the pressure to energy density ratio, $\omega=\frac{p}{\rho}$. For $\alpha \neq \pm \beta$, it is straightforward to obtain $\omega$ as
\begin{equation}
\omega = -1+\left(\alpha+\beta\right)\frac{S_2(H)-S_3(H)}{S_1(H)\beta-S_3(H)\alpha+\left(\alpha-\beta\right)\Lambda_0}.
\end{equation}
As is obvious from the above expression, the dynamical behaviour of the EoS parameter depends on the parameters of the Hubble rate $H$ and the coupling constant $\beta$. For any realistic cosmological model, the Hubble parameter is a decreasing function of time and therefore at late phase of cosmic evolution, we expect that, the functionals $S_2(H)$ and $S_3(H)$ will behave alike thereby cancelling each other at late times. Therefore, for any value of $\beta (\neq -2\pi, \neq -4\pi)$, the EoS parameter behaves as a cosmological constant ($\omega=-1$) at late epoch. However, at an early epoch, the Hubble rate assumes a very high value thereby pushes $\omega$ to a larger value.

In the limit $\beta \rightarrow 0$, the model reduces to that of GR and the EoS parameter becomes
\begin{equation}
\omega=-1+\frac{S_2(H)-S_3(H)}{\Lambda_0-S_3(H)}
\end{equation}
which becomes $\omega=-\frac{S_2(H)}{S_3(H)}$ in the absence of a cosmological constant $\Lambda_0$ term in the field equations. One can note that, similar conclusion on the dynamical evolution of $\omega$ as above may be derived for $\beta \rightarrow 0$. In other words, the dynamical behaviour of the EoS parameter will not be sensitive to the choice of the coupling constant at late times. All the trajectories of $\omega$ will behave alike at late phase of cosmic evolution. However, at an early epoch, the model will pass through different trajectories which may be $\beta$ dependent.

Another dynamical parameter is the effective cosmological constant $\Lambda_{eff}$ that appear in the equivalent Einstein Field equation for the extended gravity theory. Unlike the dynamical cosmological constant in GR, this effective cosmological constant depends on the matter field content  such as the pressure and energy density.  We can obtain $\Lambda_{eff}$ as
\begin{eqnarray}
\Lambda_{eff} &=& \left(\frac{\beta}{\alpha+\beta}\right)\left[(S_1(H)+S_3(H))-2\Lambda_0\right]+\Lambda_0.\label{eq:23}
\end{eqnarray}
In terms of the Hubble parameter, we may express $\Lambda_{eff}$ as
\begin{equation}
\Lambda_{eff} = \frac{\beta}{\alpha+\beta}\left[\frac{6}{k+2}(\dot{H}+3H^2)-2\Lambda_0 \right]+\Lambda_0. \label{eq:28}
\end{equation}
For $\alpha+\beta \neq 0$, the magnitude of the effective cosmological constant decreases with the growth of cosmic time. The sign of this quantity will depend on the sign of $\beta$. At late times, the behaviour of the effective cosmological constant depends on the contribution coming from $\dot{H}+3H^2$. In many models, this term either vanishes or have a negligible contribution. For such models,  $\Lambda_{eff}$ reduces to $\left(\frac{\alpha-\beta}{\alpha+\beta}\right)\Lambda_0$. Obviously as mentioned earlier, for a vanishing coupling constant $\beta$, it reduces to the usual time independent cosmological constant $\Lambda_0$.

\subsection{Energy Conditions}
Since energy conditions put some additional constraints on the viability of the models  we wish to calculate the different energy conditions for the constructed model in the modified gravity theory. In our formalism, the energy conditions are obtained as 
\begin{eqnarray}
\textbf{NEC}&:& \rho+p=\frac{[S_3(H)-S_2(H)]}{\alpha-\beta}\geq 0,\nonumber\\
\textbf{WEC}&:& \rho = \left(\frac{1}{\alpha^2-\beta^2}\right)\left[S_3(H)\alpha-S_1(H)\beta-\left(\alpha-\beta\right)\Lambda_0\right] \geq 0,\nonumber\\
\textbf{SEC}&:& \rho+3p=\left(\frac{1}{\alpha^2-\beta^2}\right)\left[(S_3(H)-3S_2(H))\alpha+(2S_1(H)-3S_2(H)+3S_3(H))\beta+2(\alpha-\beta)\Lambda_0\right]\geq 0, \nonumber\\
\textbf{DEC} &:& \rho -p= \left(\frac{1}{\alpha^2-\beta^2}\right)\left[(S_2(H)+S_3(H))\alpha-(2S_1(H)-S_2(H)+S_3(H))\beta-2(\alpha-\beta)\Lambda_0\right]\geq 0, \nonumber
\end{eqnarray}
where NEC, WEC, SEC and DEC respectively denote Null energy condition, Weak energy condition, Strong energy condition and Dominant energy condition. These energy conditions are expressed in terms of the Hubble parameter as
\begin{eqnarray}
\textbf{NEC} &:& \rho+p =\frac{-1}{\alpha+\beta}\left[\frac{k(k-1)}{(k+2)^2}9H^2+\frac{k+1}{(k+2)}3\dot{H}\right],\\
\textbf{WEC} &:& \rho\geq0,\\
\textbf{SEC} &:& \rho+3p =\frac{-1}{\alpha^2-\beta^2}\left[\left(\frac{3k^2+k+2}{(k+2)^2}9H^2+\frac{k+1}{(k+2)}9\dot{H}-\frac{2}{A^2}\right)\alpha\right]\nonumber\\
&&~~~~~~~~~~~-\frac{1}{\alpha^2-\beta^2}\left[\left(\frac{k^2-k-2}{(k+2)^2}27H^2+\frac{3k-1}{(k+2)}3\dot{H}-\frac{2}{A^2}\right)\beta-2(\alpha-\beta)\Lambda_0 \right]\\
\textbf{DEC} &:& \rho-p= \frac{1}{\alpha^2-\beta^2}\left[\left(\frac{k^2+3k+2}{(k+2)^2}9H^2+\frac{k+1}{(k+2)}3\dot{H}-\frac{2}{A^2}\right)\alpha\right]\nonumber\\
&&~~~~~~~~~~~+\frac{1}{\alpha^2-\beta^2}\left[\left(\frac{k^2-k-6}{(k+2)^2}9H^2+\frac{k-3}{(k+2)}3\dot{H}-\frac{2}{A^2}\right)\beta-2(\alpha-\beta)\Lambda_0 \right].
\end{eqnarray}  

We wish to compel our model in such a manner that the WEC be satisfied through out the cosmic evolution. In order to achieve this, one has to take a balance between the parameters of the Hubble rates and the choice of the coupling constant $\beta$. Since at late times, our model overlaps with $\Lambda$CDM model, the NEC and DEC are satisfied atleast at late phase of cosmic evolution. On the other hand, the SEC condition is violated at late times even though there occurs some possibility that SEC be satisfied at an early epoch. In fact, a detailed analysis on these energy condition may be possible one the cosmic dynamics is fixed up from an assumed or derived Hubble rate.

\section{Scalar field reconstruction}
In GR, the late time cosmic acceleration phenomena is modelled usually through a scalar field $\phi$ which may either be quintessence like or phantom like with the EoS parameter being $\omega \geq -1$ or $\omega \leq -1$ respectively. The action for such cases is given by
\begin{equation}
S_{\phi}= \int d^4x\sqrt{-g}\left[\frac{R}{16\pi}+\frac{\epsilon}{2} \partial_{\mu}\phi~\partial^{\mu}\phi~-V(\phi) \right],
\end{equation}
where $\epsilon=+1$ for quintessence field and $\epsilon=-1$ for phantom field. $V(\phi)$ is the self interacting potential of the scalar field. The scalar field dynamically rolls down the potential and thereby mediating for cosmic acceleration. In this work, we wish to draw a correspondence between the geometrically modified gravity theory discussed above with that of the scalar field cosmology and also wish to reconstruct the scalar field along with the scalar potential. In a flat Friedman background, the energy density and pressure are expressed by
\begin{eqnarray}
\rho_{\phi} &=& \frac{\epsilon}{2}\dot{\phi}^2+V(\phi),\\
p_{\phi} &=& \frac{\epsilon}{2}\dot{\phi}^2-V(\phi).
\end{eqnarray}

A direct correspondence of our model with the scalar field yields
\begin{eqnarray}
\dot{\phi}^2 &=& \frac{\epsilon}{\alpha-\beta}\left[S_3(H)-S_2(H)\right],\\
V(\phi) &=& \left(\frac{1}{\alpha^2-\beta^2}\right)\left[\left\{S_2(H)+S_3(H)\right\}\alpha-\left\{2S_1(H)-S_2(H)+S_3(H)\right\}\beta-2(\alpha-\beta)\Lambda_0\right].
\end{eqnarray}
Since the factor $S_3(H)-S_2(H)$ decreases with the cosmic evolution, we expect the magnitude of $\dot{\phi}$ to decrease with cosmic time. It is worth to mention here that the exact behaviour of the scalar field will be model dependent and can be investigated with some specific evolutionary behaviour of the Hubble rate.

\begin{figure}[ht!]
\begin{center}
\includegraphics[width=0.85\textwidth]{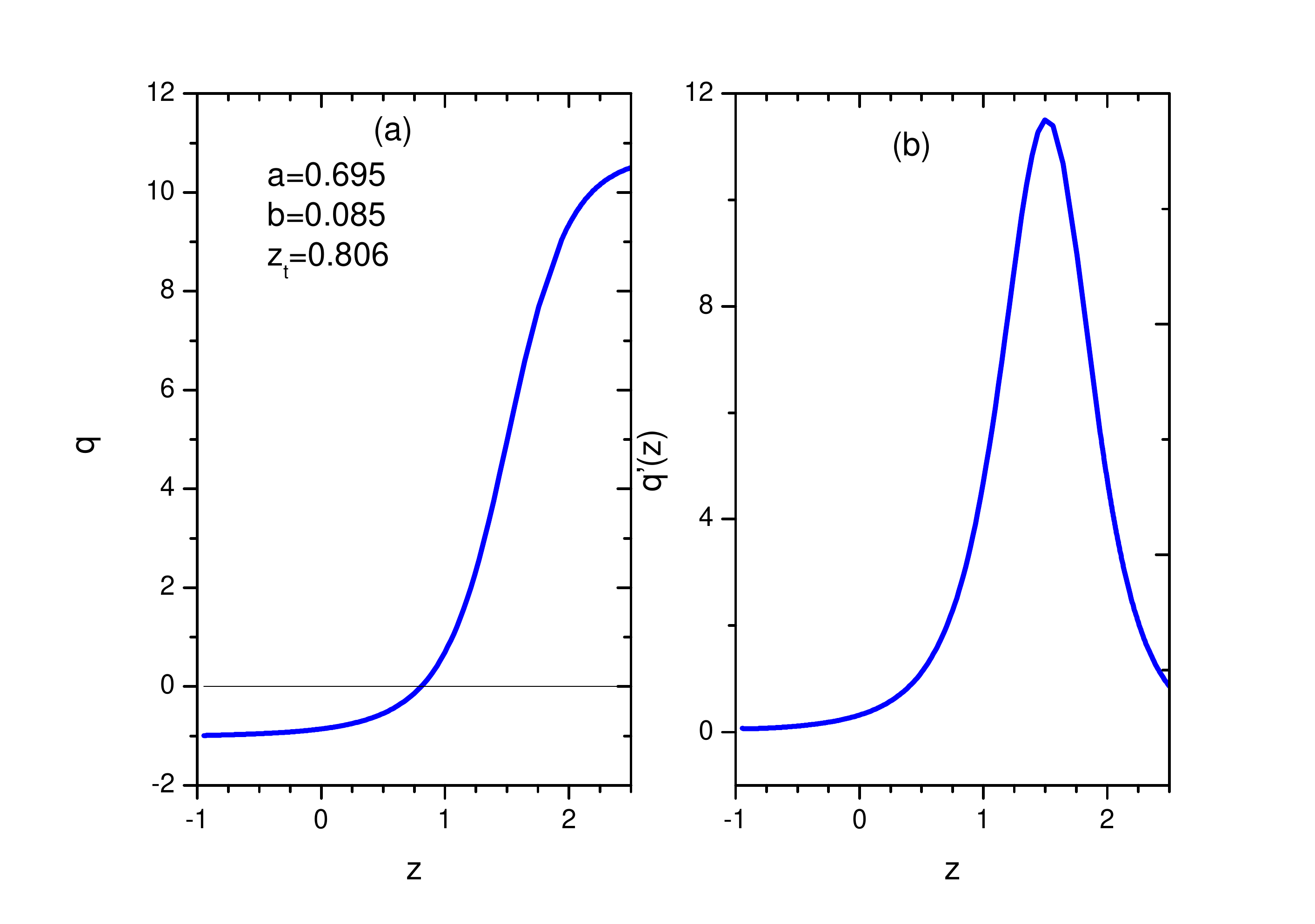}
\caption{(a)Deceleration parameter for HSF showing the transition redshift (b) $q^{\prime}(z)$ as a function of redshift. The model does not favour a slowing down at late phase.}
\end{center}
\end{figure}

\section{Model with a Hybrid Scale Factor}
The formalism developed in this work can be used to investigate certain aspects of cosmic dynamics. One can note that all the dynamical properties are expressed in terms of the Hubble rate $H$. Therefore, for a given dynamics, if the Hubble rate is known, then it becomes easy to track the evolution history. In view of this, we employ a hybrid scale factor (HSF) $\mathcal{R}=e^{at}t^b$ in the formalism. Here $a$ and $b$ are the model parameters and are constrained from different observational and physical basis. The reason behind the choice of such a scale factor is that it simulates a transition from a decelerated universe in recent past to an accelerated one. Moreover, the dynamical behaviour of HSF as predicted remains intermediate to that of the power law expansion and exponential expansion. Transit redshift $z_t$ is an important cosmological parameter which has been recently constrained from an analysis of type Ia Supernova observation and Hubble parameter measurements  as $z_t=0.806$ \citep{Jesus17, Farooq17}. In a recent work, we have constrained the parameters of HSF as $a=0.695$ and $b=0.085$ so as to obtain a transition redshift $z_t=0.806$ \citep{Mishra18a}. The Hubble parameter for the HSF is given by $H=a+\frac{b}{t}$ so that the directional Hubble rates become $H_1=\frac{3k}{k+2}\left(a+\frac{a}{b}\right)$ and $H_2=H_3=\frac{3}{k+2}\left(a+\frac{a}{b}\right)$. The deceleration parameter for HSF is $q=-1+\frac{b}{(at+b)^2}$. In Fig. 1(a), we have shown the deceleration parameter $q$ which displays the signature flipping behaviour at a suitable transit redshift. The deceleration parameter decreases from $q=-1+\frac{1}{b}\simeq 10.765$ at an early time to $q \simeq -1$ at late time. At the present epoch, this models predicts the deceleration parameter to be $q=-0.86$. Recently, analysis from a host of Hubble parameter measurements and type Ia Supernova observational data casts a doubt that, the universe has already reached the peak of its acceleration and may be we are currently witnessing a possible slowing down \citep{Shaef09,Zhang18}. Such a feature is investigated through the reconstruction of the slope of the deceleration parameter from observations. In order to check whether the HSF can predict such a feature we have plotted the function $q^{\prime}(z)=\frac{dq}{dz}$ as a function of redshift $z=\frac{\mathcal{R}_0}{\mathcal{R}}-1$ in Fig.1(b). Here $\mathcal{R}_0$ is the scale factor at present epoch. The figure shows that there is no slowing down in cosmic acceleration at late phase of cosmic time.  However, we find an interesting feature where $q^{\prime}(z)$ peaks up at around $z=1.5$. In order to have a quantitative idea about the deceleration parameter, we have listed some of its values at different epochs in table-I.

\begin{figure}[ht!]
\begin{center}
\includegraphics[width=0.85\textwidth]{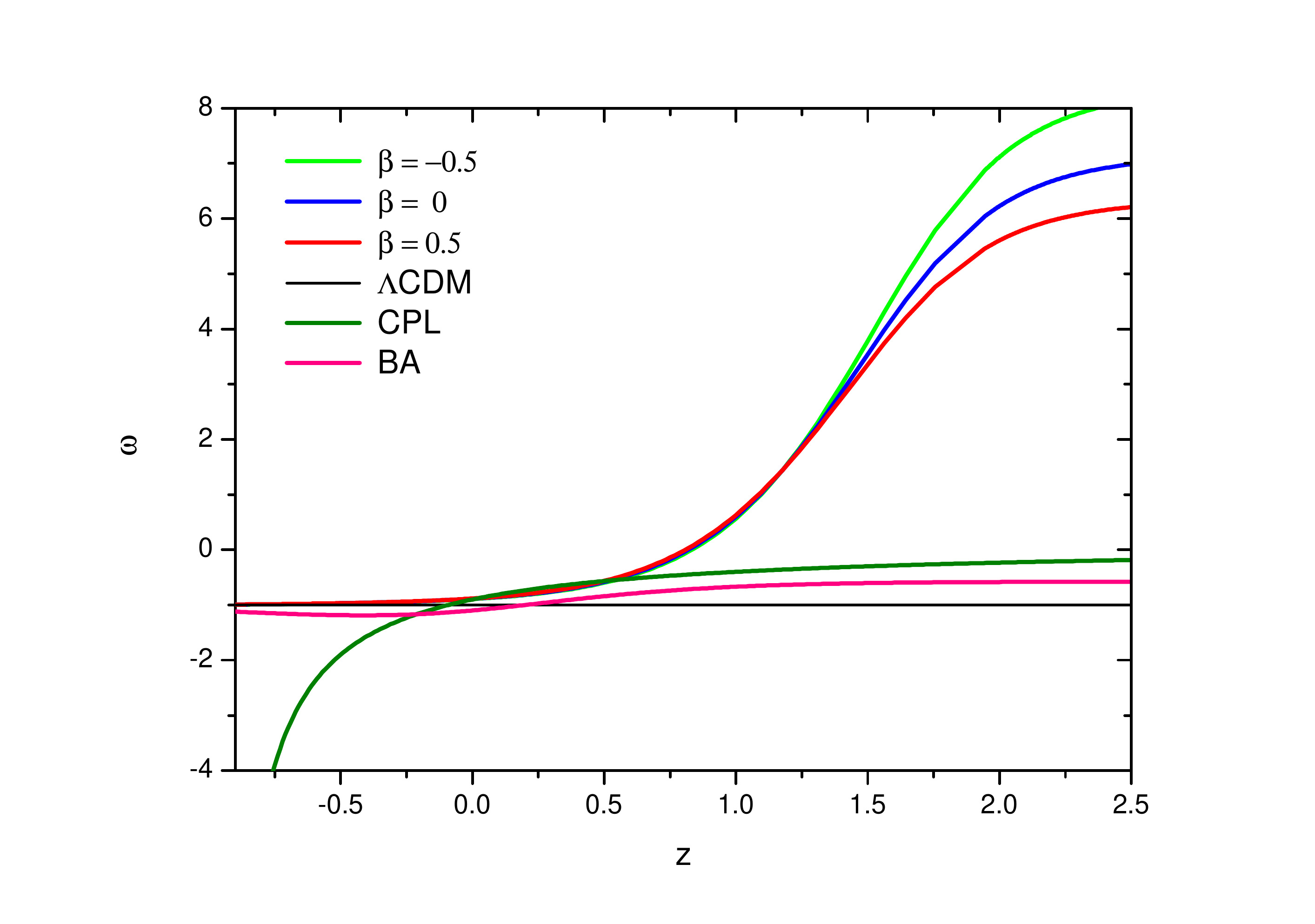}
\caption{ Equation of State parameter for three representative values of the coupling constant $\beta$. $k=1.0000814$}
\end{center}
\end{figure}

\begin{table}[ht]
\caption{Deceleration parameter at different epochs} 
\centering 
\begin{tabular}{c|c| c } 
\hline\hline 
 epoch         & z       &  $q$ \\ [0.5ex] 
\hline 
Late phase	   &  $-0.9$ &  -0.99   \\
Present        &  $0$    &  -0.86    \\
At transit     &  $0.8$  &       0    \\ 
Early phase	   &  1.5	 &   4.93       \\
\hline 
\end{tabular}
\label{table:nonlin} 
\end{table}

\begin{figure}[ht!]
\begin{center}
\includegraphics[width=0.85\textwidth]{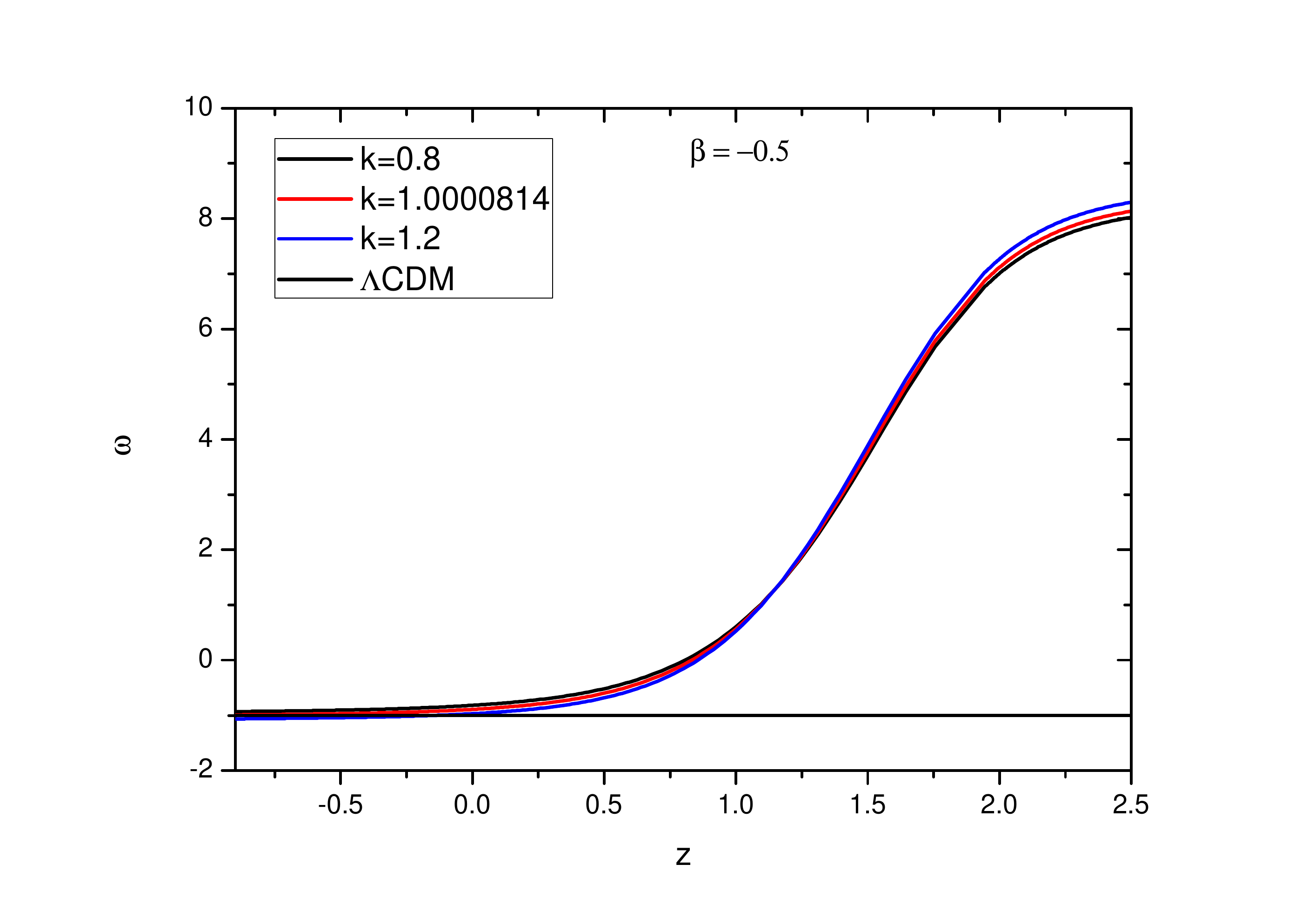}
\caption{Equation of State parameter for three representative values of the anisotropy parameter $k$. $\beta=-0.5$}
\end{center}
\end{figure}
\begin{figure}[ht!]
\begin{center}
\includegraphics[width=0.85\textwidth]{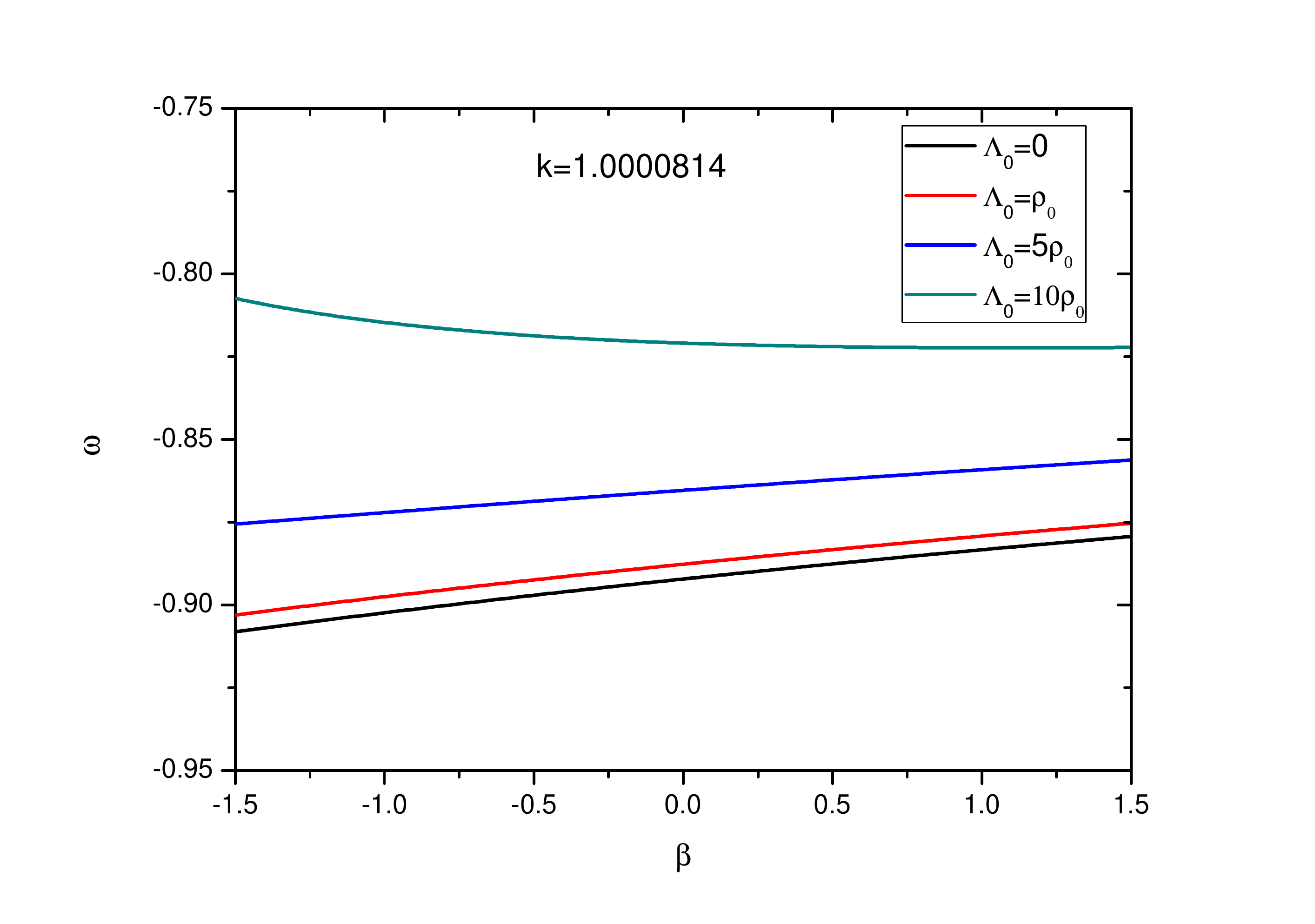}
\caption{Equation of State parameter as function of coupling constant $\beta$ for three representative values of the cosmological constant . $k=1.0000814$}
\end{center}
\end{figure}

\begin{figure}[ht!]
\begin{center}
\includegraphics[width=0.85\textwidth]{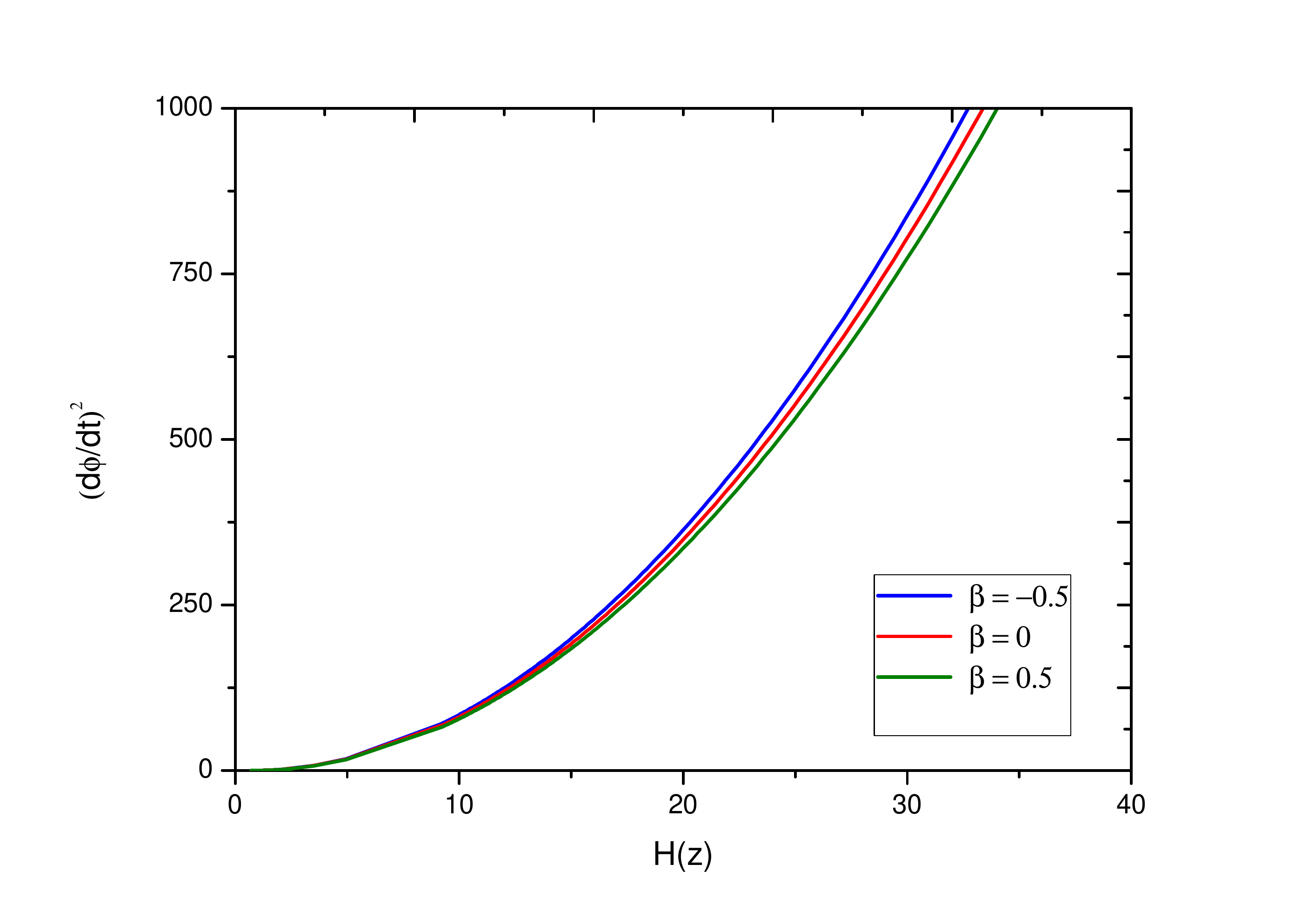}
\caption{Squared Slope of Reconstructed Scalar field as a function of Hubble rate.}
\end{center}
\end{figure}

\begin{figure}[ht!]
\begin{center}
\includegraphics[width=0.85\textwidth]{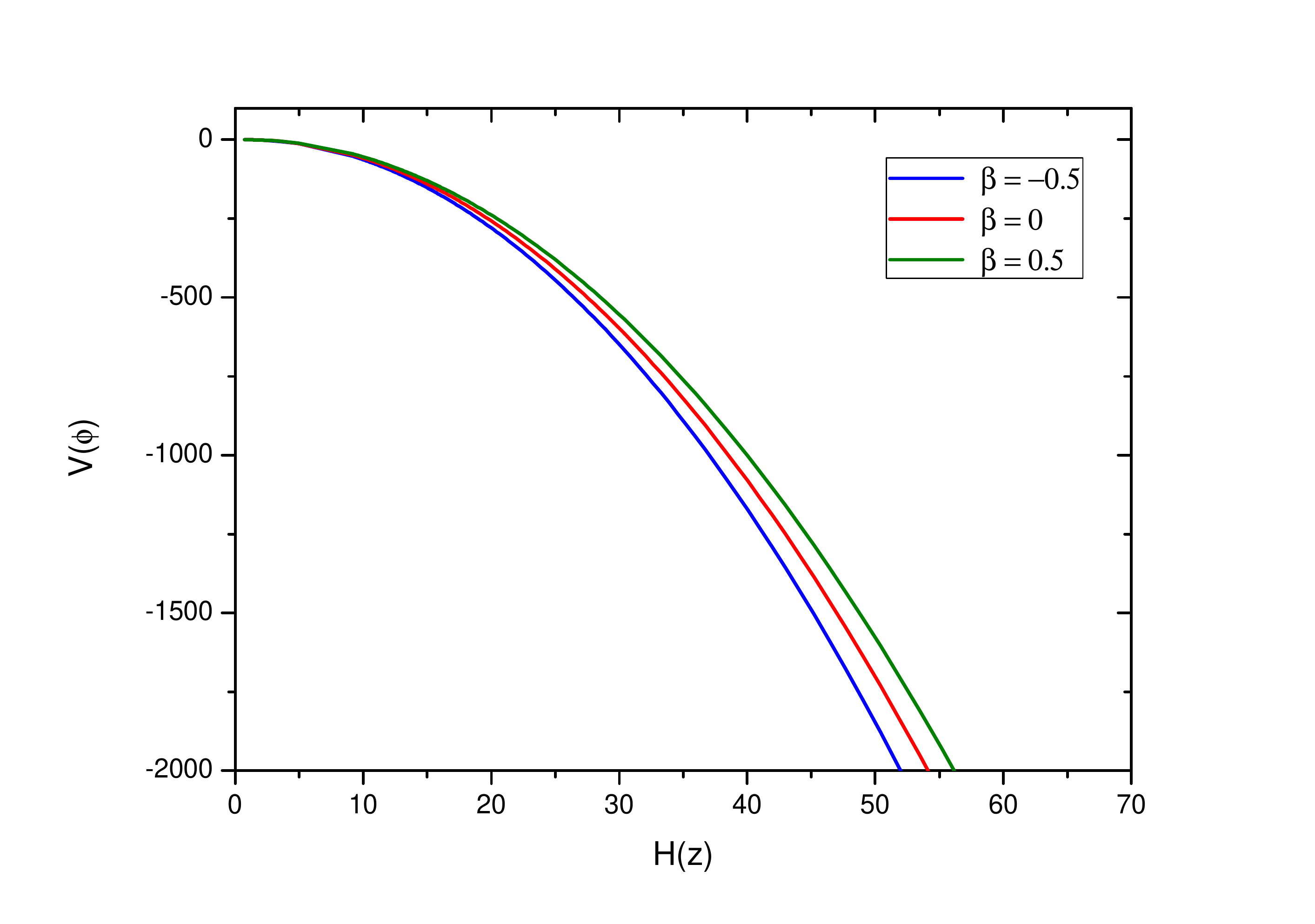}
\caption{Scalar potential as a function of Hubble rate.}
\end{center}
\end{figure}
In Fig.2, the dynamical aspect of the model is assessed through the plot of the EoS parameter as function of redshift. In the figure $\omega$ is shown for a fixed anisotropic parameter $k=1.0000814$ and for three different values of the coupling constant $\beta$. In general, the EoS parameter decreases from an initial positive value to behave like a cosmological constant at late phase. The initial positive value depends on the choice of $\beta$. As expected, at late times, the model is insensitive to the choice of the coupling constant $\beta$. However at an early epoch, the EoS parameter evolves through different trajectories for different choices of $\beta$. Lower is the value of $\beta$, higher is the slope of the EoS curve at late times. In other words, at an early epoch, the $\omega$ trajectory for low values of $\beta$ remains in the top of others. The $\beta$ dependent splitting in these trajectories is visible at around $z=1.5$. At the present epoch, the model predicts an EoS of $\omega=-0.89$ which is well within the observational constraints. In the figure, for a comparison, we have also shown the trajectories for two well known $\omega$ parametrizations such as CPL \citep{CPL01,Linder03} and BA \citep{BA12}  given by  $\omega(z)=\omega_0+\omega_a\frac{z}{1+z}$ and $\omega(z)=\omega_0+\omega_0\frac{z(1+z)}{1+z^2}$ respectively. It is clear from the comparison that, in a time zone in the range $-0.25 \leq z \leq 0.5$, EoS from HSF is in close agreement with other models. 

The effect of the anisotropic parameter $k$ on the EoS is investigated in Fig.3. In the figure we have shown the evolution of $\omega$ for a given coupling constant $\beta=-0.5$ for three representative values of $k$ namely $k=0.8, 1.0000814$ and $1.2$. We note here that, we have considered a specific shear expansion to Hubble rate ratio in the present epoch within the observational limits and have constrained $k$ to be $1.0000814$. This value of $k$ shows a very little departure from its isotropic value. Unlike that of the coupling constant, cosmic expansion anisotropy affects the cosmic dynamics both at an early epoch and at late times. The effect of cosmic anisotropy is almost symmetrical about $z=1.25$. At epochs $z < 1.25$, higher the value of $k$, lower is the value of $\omega$ and at epochs $z\geq 1.25$, the EoS shows an opposite behaviour i.e higher the value of $k$, higher is the value of  $\omega$. A quantitative idea on the effect of the cosmic expansion anisotropy on the EoS can be obtained from the values listed in Table.II.

\begin{table}[ht]
\caption{Variation of EoS parameter with anisotropy parameter} 
\centering 
\begin{tabular}{c|c| c |c|c} 
\hline\hline 
 epoch         & z       &  $k=0.8$ &  $k=1.0000814$ & $k=1.2$ \\ [0.5ex] 
\hline 
Late phase	   &  $-0.9$ &  -0.932  &  -0.994        &   -1.064 \\
Present        &  $0$    &  -0.819  &  -0.892        &   -0.971  \\
At transit     &  $0.8$  &  -0.012  &  -0.08         &   -0.152   \\ 
Early phase	   &  1.5	 &   3.68   &   3.76		 &    3.86     \\
\hline 
\end{tabular}
\label{table:nonlin1} 
\end{table}

\begin{table}[ht]
\caption{ EoS parameter at present epoch for different values of cosmological constant $\Lambda_0$} 
\centering 
\begin{tabular}{c|c|c|c|c|c} 
\hline\hline 
$\Lambda_0$   &  $\omega_0 (\beta=-1)$    & $\omega_0 (\beta=-0.5)$ & $\omega_0 (\beta=0)$ & $\omega_0 (\beta=0.5)$& $\omega_0 (\beta=1)$     \\ [0.5ex] 
\hline 
0	        & -0.902 &-0.897 &  -0.892& -0.888 & -0.883\\
$\rho_0$    & -0.898&-0.892 &  -0.888& -0.883& -0.879\\
$5\rho_0$    &-0.872&-0.869 &  -0.866&-0.862 & -0.859 \\ 
$10\rho_0$   &-0.815&-0.819 &  -0.821& -0.822&  -0.822  \\
\hline 
\end{tabular}
\label{table:nonlin2} 
\end{table}
\pagebreak

In order to assess a simultaneous effect of the coupling constant $\beta$ and the cosmological constant $\Lambda_0$ on the EoS parameter, we have shown the variation of $\omega$ at the present epoch with respect  to $\beta$ for four different values of $\Lambda_0$ and a given cosmic anisotropy $k=1.0000814$ in Fig.4. The representative values of $\Lambda_0$ are considered as multiples of the present value of energy density $\rho_0$ as calculated from the present model. $\omega$ increases with the increase in the coupling constant for a given value of $\Lambda_0$. Also for a given $\beta$, it increases with the  increase in $\Lambda_0$. However, the net variation of $\omega$ with respect to $\beta$ decreases with an increase in  $\Lambda_0$. To get a quantitative view, in table-III, the values of the EoS parameter at the present epoch are given for some representative values of the cosmological constant and the coupling constant.

In Fig.5, the quintessence like scalar fields are reconstructed from our model for three representative values of the coupling constant $\beta$. The anisotropy parameter $k$ and the time independent cosmological constant $\Lambda_0$ are chosen to be $1.0000814$ and $\rho_0$ respectively. As per expectation, the scalar field is found to decrease with the cosmic expansion. In Fig. 6, the evolution of the self interacting potential for the quintessence like scalar field is plotted. The self interacting potential increases with cosmic expansion. The choice of the coupling constant $\beta$ does not affect the general evolutionary behaviour of these two quantities. However with an increase in $\beta$ value at a given epoch, the scalar field decreases and the potential increases.

\section{Diagnostic approach}
There are two important diagnostic approaches used in literature. They are the determination of the state finder pair $\{j,s\}$ in the $j-s$ plane and the $Om(z)$ diagnostics. These geometrical diagnostic approaches are useful tools to distinguish different Dark Energy models. While the state finder pair involve third derivatives of the scale factor, the $Om(z)$ parameter involve only the first derivative of the scale factor appearing through the Hubble rate $H(z)$.

\subsection*{Statefinder pair}
State finder pairs provide an useful tool to distinguish Dark Energy models since they involve the third derivative of the scale factor. They are defined as 

\begin{eqnarray} 
 j &=&  \frac{\dddot{\mathcal{R}}}{\mathcal{R}H^3}=\frac{\ddot{H}}{H^3}-(2+3q),\\
 s &=& \frac{j-1}{3(q-0.5)}.   
\end{eqnarray}
 
 In our formalism, the deceleration parameter is a time varying quantity and therefore the state finder pair evolve with time. In Fig.7, the $j-s$ trajectory in the $j-s$ plane is shown for the HSF considered in this work. We observe that, our model evolves to overlap with the $\Lambda$CDM model.

\begin{figure}[ht!]
\begin{center}
\includegraphics[width=0.85\textwidth]{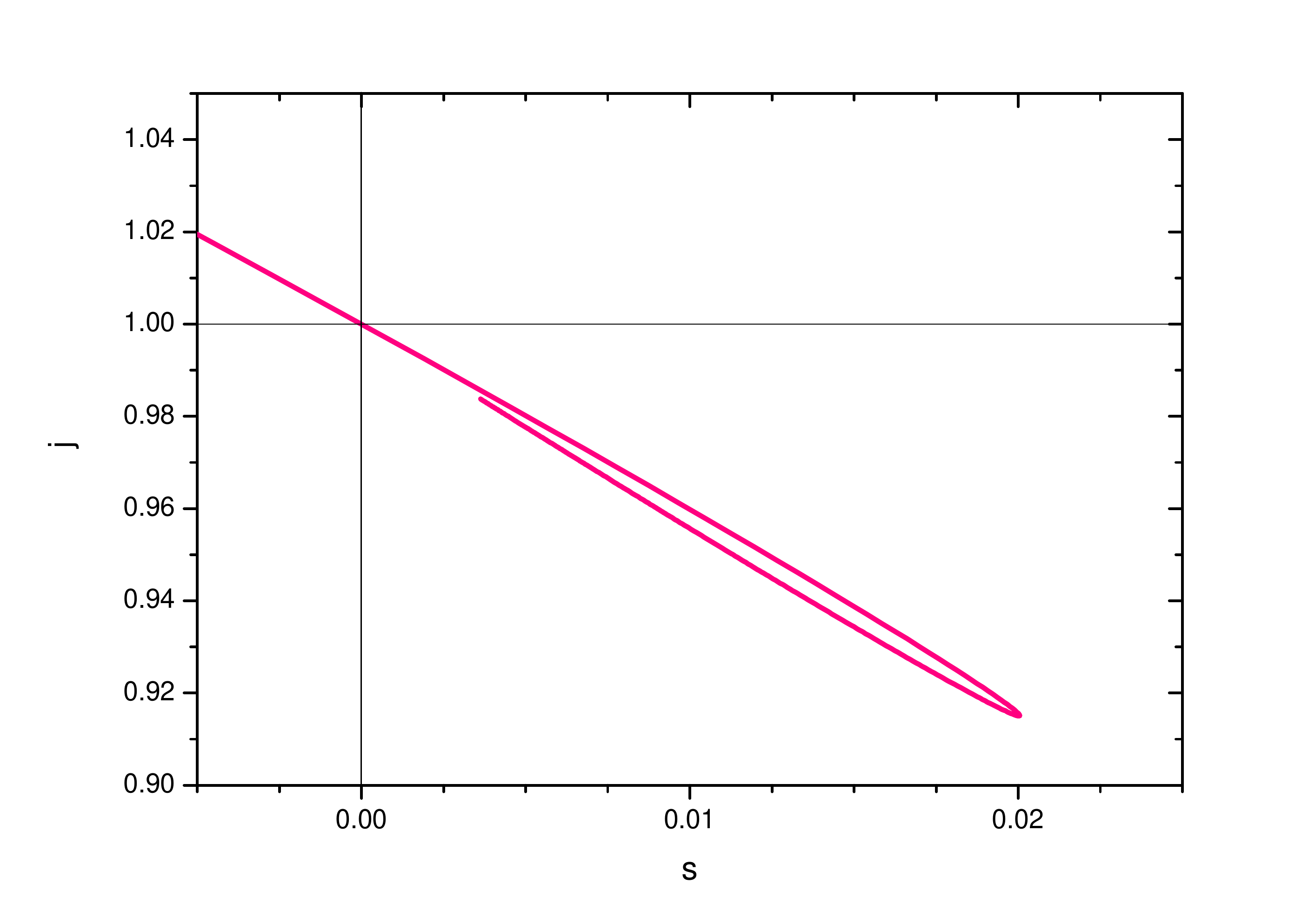}
\caption{$j-s$ trajectory in the $j-s$ plane.}
\end{center}
\end{figure}

\subsection*{$Om(z)$ diagnostic}
Another geometric diagnostic methods is the $Om(z)$ diagnostic that involves first derivative of the scale factor and therefore becomes easier to apply to distinguish between different Dark Energy models \citep{Sahni08}. The $Om(z)$ parameter is defined by

\begin{equation}
Om(z) = \frac{E^2(z)-1}{(1+z)^3-1},
\end{equation}
where $E(z)=\frac{H(z)}{H_0}$ is the dimensionless Hubble parameter. Here $H_0$ is the Hubble rate at the present epoch. If $Om(z)$ becomes a constant quantity, the DE model is considered to be a cosmological constant model with $\omega = -1$. If this parameter increases with $z$ with a positive slope, the model can be a phantom model with $\omega < -1$. For a decreasing $Om(z)$ with negative slope, quintessence model are obtained ($\omega >-1$). In Fig.8, the $Om(z)$ parameter for HSF is shown as a function of redshift. It can be observed from the figure that, the model looks like a cosmological constant model for a substantial time zone in the recent past ($0\leq z \leq 0.7$). Before this period, the model evolves as a quintessence field.
\begin{figure}[ht!]
\begin{center}
\includegraphics[width=0.85\textwidth]{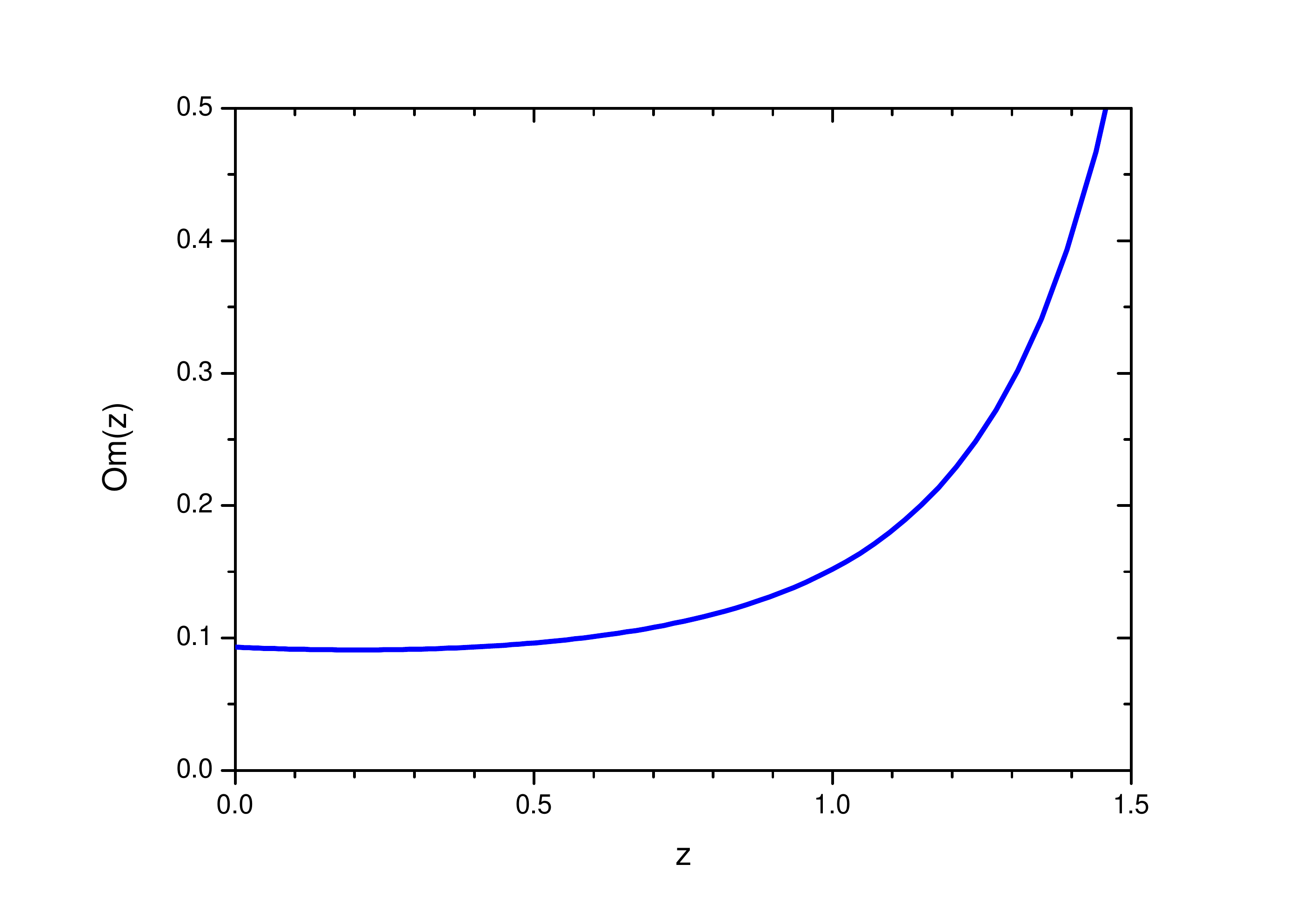}
\caption{$Om(z)$ parameter.}
\end{center}
\end{figure}

\section{Conclusion}
In the present work, we have constructed a cosmological model in an extended theory of gravity by considering the functional $f(R,T)= R+2\Lambda_0+\beta T$, where $\Lambda_0$ is a constant. This model reduces to the usual GR equations with a cosmological constant in the limit of a vanishing coupling constant $\beta$.  Investigation of dynamical features of universe in such an extended theory requires an involved calculation. In order to study certain dynamical cosmic aspects, we have adopted an interesting approach in the present work and obtained the expressions in a more general manner. Although the cosmological principle assuming a homogeneous and isotropic universe is a good approximation to the present universe, it is yet to be proven in high energy scales. In view of this, we have considered an anisotropic universe which is more general than the FRW model for our purpose. The anisotropic model we have constructed can be applicable to any amount of cosmic anisotropy. The anisotropic behaviour can be assessed through the value of the anisotropy parameter at the present epoch which has been constrained as $k=1.0000814$. This value of cosmic anisotropy leads to $\left(\frac{\sigma}{H}\right)_0 = 4.7 \times 10^{-5}$. The expansion asymmetry from our model is obtained to be $\frac{\triangle H}{H}=0.814\times10^{-4}$ which is in conformity with the observations. 

A dynamically changing universe with a feature of early deceleration and late time cosmic acceleration is simulated through a hybrid scale factor. The parameters of the HSF are constrained from some physical basis to reproduce the transition redshift as obtained from different observational analysis. This HSF provides a good estimate of the deceleration parameter  and the Hubble rate at the present epoch. Recently there has been a belief that, we are at the peak of the cosmic acceleration and the universe is now slowing down. We have investigated such a feature of the universe employing the HSF and obtained that there is no such slowing down in recent past or recent future.
 
The dynamical behaviour of the model is assessed through the calculation of the EoS parameter employing the HSF. The EoS parameter decreases from a positive value in an early phase to a value closer to $-1$ at late times. The behaviour of the EoS parameter is sensitive to the choice of the coupling constant at late times and  all the trajectories of EoS parameter for different choices of the coupling parameter behave alike at late phase. However at an early phase, the trajectory splits into different $\beta$ channels. Trajectory with low values of $\beta$ lies in the top of all trajectories. Different diagnostic approaches have been adopted to analyse the viability of the present constructed model. At late phase, the model looks like a $\Lambda CDM$ model for a substantial cosmic time zone. In the rest phase, it behaves as a quintessence field. 

\section*{Acknowledgement}
BM and SKT thank IUCAA, Pune (India) for hospitality and support during an academic visit where a part of this work is accomplished. BM and ST acknowledge DST, New Delhi, India for providing facilities through DST-FIST lab, Department of Mathematics, where a part of this work was done. ST thanks University Grants Commission (UGC),New Delhi, India, for the financial support to carry out the research work.

\end{document}